%
%
%
\magnification=1200

\hyphenation{Chan-d-ra-sek-har Schwarz-schild an-is-o-tro-pic Max-well
An-to-nov Max-well-ian Hertz-sprung Hertz-sprung-Rus-sell
nec-es-sa-ri-ly nec-es-sa-ry nec-es-sa-ries ne-ces-si-tate ne-ces-si-tates
ne-ces-si-tated ne-ces-si-tating ne-ces-si-tous ne-ces-si-ty ne-ces-si-ties
arc-min-ute arc-min-utes}

\def\iso#1#2{\hbox{${}^{#1}{\rm #2}$}}
\def\markcite#1{#1}
\def\Menv{M_{\rm env}}     

\hoffset=0 true in
\hsize=6.75 true in
\voffset=0 true in
\vsize=9 true in
\overfullrule=0pt

%
\baselineskip=14 pt plus 1pt minus 1pt
%
%
$  $
\vskip 0.5truein
\centerline{{\bf Cosmological Implications of $\bf^3$He Destruction
 on the Red Giant Branch}}
\bigskip
\centerline{Arnold I. Boothroyd\footnote{$^1$}{Now at Dept.\ of Mathematics,
Monash University, Clayton, VIC~3168, Australia} and Robert A. Malaney}
\bigskip
\centerline{{\it Canadian Institute for Theoretical Astrophysics,}}
\centerline{{\it University of Toronto, Toronto, Ontario, CANADA\ \ M5S 1A7}}
\centerline{Email: boothroy@vulcan.maths.monash.edu.au ,
 malaney@cita.utoronto.ca}

\vskip 0.5 true in

\centerline{ABSTRACT}
\medskip

Observations of stellar CNO isotope ratios indicate the
presence of additional mixing processes on the red giant branch.
An estimate of the resulting stellar $^3$He depletion is made, as
a function of stellar mass and metallicity.
Based on stellar nucleosynthesis and galactic chemical evolution
calculations, we determine the degree to which the destruction of $^3$He
associated with such extra mixing processes can influence the
inferred primordial abundance of D$+^{3}$He. We find that the extra
mixing processes may increase the upper limit
of the inferred primordial (D$+^{3}$He)/H ratio by $\sim 20\%$.
The implications of this
for baryonic dark matter bounds, and constraints on the
relativistic degrees of freedom in the early universe, are discussed.

\vskip 0.5 true in
\leftline{\hbox{{\it Subject Headings\/}: }early universe ---
 Galaxy: abundances --- Galaxy: evolution ---}
\leftline{$\phantom{\hbox{{\it Subject Headings\/}: }}$stars: abundances ---
 stars: AGB and Post-AGB --- stars: giants}

\vskip 1 true in
\centerline{{\it Preprint\/} {\bf CITA-95-27}}
\bigskip
\centerline{{\it Submitted to Astrophysical Journal, December 1995}}

\vfil\eject

\medskip
\centerline{1. INTRODUCTION}
\medskip

The upper bound on the inferred primordial abundance of D$+^{3}$He
plays an important role in cosmology and on the nature of the dark matter.
Coupled with calculations of standard big bang nucleosynthesis (SBBN)
(see, e.g., \markcite{Yang {\it et al.}\ 1984};
\markcite{Walker {\it et al.}\ 1991};
\markcite{Smith {\it et al.}\ 1993}; \markcite{Kernan \& Krauss 1994};
\markcite{Copi {\it et al.}\ 1995a}), this bound allows for a
lower limit on $\Omega_b$, the baryonic-to-critical density ratio
of the universe.  Knowledge of the lower limit to $\Omega_b$ then allows
an upper limit on the relativistic degrees of freedom
in the early universe to be set.

Stars burn their initial deuterium to \iso3{He} while still on the pre-main
sequence.  Low mass stars ($1 - 3\>M_\odot$) create a \iso3{He}-rich pocket
in their interior during main sequence burning
(\markcite{Iben 1967}; \markcite{Sackmann \& Boothroyd 1995a}); this pocket
will subsequently be dredged up to the surface on the the red giant branch
(RGB), and injected into the interstellar medium via stellar mass loss on
the RGB and on the asymptotic giant branch (AGB).  Thus the sum
of D+\iso3{He} in the interstellar medium was predicted
to increase with time, due to this stellar processing.
However, \markcite{Rood, Bania, \& Wilson (1984)}
pointed out that the apparent observed trend with galactocentric radius
of the \iso3{He} abundance was the opposite of what one would expect if
stellar processing was increasing
the galactic \iso3{He} abundance.  Recent observations of a high ratio
$\rm D/H \approx 2.5 \times 10^{-4}$ in a high-redshift quasar absorber
(\markcite{Songaila {\it et al.}\ 1994};
\markcite{Carswell {\it et al.}\ 1994}) has
sparked new interest in the subject, since it implied a much larger
primordial (D+\iso3{He})/H value than had been inferred from the galactic
observations (provided that the observed absorption line did indeed
correspond to deuterium and not some coincidental interloper cloud
happening to lie at the corresponding velocity; note also that in a
different line of sight \markcite{Tytler \& Fan [1994]} more recently
reported a preliminary value as low as $\rm D/H \approx 2 \times 10^{-5}$
in a high-redshift quasar absorber, and that
\markcite{Levshakov \& Takahara [1995]} have suggested that
problems with the turbulent modeling of the absorbing systems could
result in very large errors in the derived D~abundance.)

The high D/H observation of \markcite{Songaila {\it et al.}\ (1994)} and
\markcite{Carswell {\it et al.}\ (1994)}
implied a decline by a factor of order~6 from the
primordial (D+\iso3{He})/H value to the presolar value.
\markcite{Galli {\it et al.}\ (1994, 1995)}
suggested that a low-energy resonance in the $\iso3{He}+\iso3{He}$
reaction could increase this rate sufficiently for low mass stars to
become net destroyers of \iso3{He}.  \markcite{Hogan (1995)} suggested a less
speculative mechanism, namely, {\it deep extra mixing\/} below the (relatively
cool) base of the conventional convective envelope on the RGB,
resulting in nuclear processing of envelope material subsequent to
first dredge-up, i.e., ``{\it cool bottom processing\/}'' (CBP).  Such
extra mixing and resulting CBP is needed in order to understand the puzzle
of the low \iso{12}C/\iso{13}C ratios observed in low mass Population~I RGB
stars, and the \iso{12}C depletion and the [O/Fe] -- [Na/Fe] anticorrelation
discovered in low mass Population~II RGB stars (see, e.g.,
\markcite{Dearborn, Eggleton, \& Schramm 1976};
\markcite{Genova \& Schatzman 1979}; \markcite{Gilroy 1989};
\markcite{Gilroy \& Brown 1991}; \markcite{Dearborn 1992};
\markcite{Smith \& Tout 1992}; \markcite{Charbonnel 1994};
\markcite{Boothroyd, Sackmann, \& Wasserburg 1995}, hereafter BSW95;
\markcite{Wasserburg, Boothroyd, \& Sackmann 1995}, hereafter WBS95;
\markcite{Denissenkov \& Weiss 1995}; \markcite{Charbonnel 1995}).
Recently, \markcite{BSW95}
and \markcite{WBS95} pointed out that such deep extra
mixing should also occur in AGB stars of low mass, in order to account
for the anomalously low \iso{18}O abundances observed in these stars.
(Note that the structure of a star on the AGB is very similar to that
of a low-mass star on the RGB; the main difference is that there is a
helium-exhausted degenerate core inside the hydrogen-exhausted core.
Since the extra mixing and CBP take place outside the core, one would
expect them to take place on the AGB as well as on the RGB\hbox{}.)
That $^3$He can be destroyed in low mass stars raises the serious possibility
that the inferred primordial abundance of D$+^{3}$He could be increased,
and consequently the lower bound of $\Omega_b$ reduced.

It is the purpose of this report to investigate this interesting possibility
in some detail.  In section~II we present our calculations
of the $^3$He survival fractions in stars, based on nucleosynthesis
models normalized to the observed
$^{12}$C/$^{13}$C and $^{16}$O/$^{18}$O ratios.  Note that \markcite{WBS95}
considered only a $1 \>M_\odot$ star of solar metallicity;
\markcite{Sackmann \& Boothroyd (1995a)}
used this to make a rough estimate of the average \iso3{He}
survival fraction~``$g_3$'' for solar metallicity stars.  The present work
provides improved estimates as a function of both mass and metallicity.
In section~III
we couple these calculations to models of galactic chemical
evolution in order to ascertain the highest value of the primordial
D$+^{3}$He abundance consistent with its presolar value.
Our basic conclusions will be that additional RGB mixing allows for a
consistent picture of D$+^{3}$He chemical evolution, and that
the primordial D$+^{3}$He abundance
remains as the strongest lower bound on $\Omega_b$.

\bigskip
\centerline{2. STELLAR ${}^3$He PROCESSING}
\medskip

\markcite{Sackmann \& Boothroyd (1995a)} present the results of
first and second dredge-up (occurring prior to CBP)
on the \iso3{He} abundance in stellar envelopes as a function of mass
and metallicity, from standard stellar evolution models.  In these models,
in the absence of any definitive model of \iso3{He} evolution with
metallicity, the initial stellar \iso3{He} abundance was chosen by setting
$\iso3{He}/Y = 3 \times 10^{-4}$ (abundance mass fraction), where the
\iso4{He} mass faction~$Y$ varied according to $Y = 0.24 + 2 Z$ with
the stellar metallicity~$Z$.  Reasonable variations in the initial
stellar \iso3{He} abundances would have relatively little effect on
the conclusions of this paper.

For CBP models, parametric computations were performed,
with envelope structures obtained from models of a solar-metallicity
$1\>M_\odot$ star (\markcite{Sackmann, Boothroyd, \& Kraemer 1993}),
not long after first dredge-up on the RGB,
and prior to the first He-shell flash (thermal pulse) on the
AGB, as discussed in \markcite{WBS95}.
In our two-stream ``conveyor-belt'' circulation model, matter from the
bottom of envelope convection streamed downward,
reaching a maximum temperature~$T_P$,
then returned upward and was mixed with the convective envelope (i.e., a
composition advection equation with nuclear burning, and no mixing
between downward and upward streams).  Envelope and
stream compositions were followed through time.
The value of~$T_P$ was treated as a free parameter;
values selected for discussion were those satisfying the observational data,
as discussed in \markcite{WBS95}.  The results of these $1\>M_\odot$ models
for \iso3{He} (and other light elements) are presented in
\markcite{Sackmann \& Boothroyd (1995a)}.  These models suggested that
low mass stars should deplete \iso3{He} (rather than create it), but
gave little indication of the \iso3{He} depletion expected in stars
as a function of mass and metallicity.  Note that \markcite{Charbonnel (1995)}
has modelled CBP in a low-mass Population~II star, using a diffusion algorithm,
and also finds a large reduction in the envelope \iso3{He} abundance.

\markcite{Boothroyd \& Sackmann (1995)} were able to make a rough
{\it relative\/} estimate of the amount of processing of the CNO isotopes
due to CBP, as a function of stellar mass and metallicity.  The amount
of hydrogen burned on the RGB or AGB is proportional to the advance of
the core mass~$\Delta M_c$, envelope dilution leads to a
factor~$(\Menv)^{-1}$, and variation of the metallicity changes the
temperature of CNO burning in the shell, leading to a
factor~$(Z_{\rm CNO})^{-1}$, i.e., roughly~$Z^{-1}$.  Models of stellar
evolution allow one to estimate the value of~$\Delta M_c$, and the resulting
estimate was presented in \markcite{Boothroyd \& Sackmann (1995)}.

The nuclear rates for the destruction of~\iso3{He} are less
temperature-sensitive than the CNO-burning rates, and thus a relative
estimate of \iso3{He}-destruction should have a factor of
roughly~$Z^{-0.8}$, rather than the above factor of~$Z^{-1}$.  Thus the
CNO-processing estimate of \markcite{Boothroyd \& Sackmann (1995)} was
multiplied by $Z^{0.2}$ to get a relative estimate of \iso3{He}-destruction
due to CBP\hbox{}.  This relative estimate was normalized by
the $Z = 0.02$, $1\>M_\odot$ parametric CBP models.
The amount of \iso3{He}-destruction in these models was determined on
the RGB by the constraint that \iso{13}C-production must match the
observations of \iso{12}C/\iso{13}C.  On the AGB, there are two possible
normalizations.  One may make the conservative assumption that the
normalization should be the same as for the RGB\hbox{}.  Alternatively,
one may normalize by attempting to match the observed \iso{18}O-depletion
on the AGB (although these observations are less accurate than the RGB
carbon isotope observations).  The latter normalization results in much
more AGB \iso3{He}-destruction than the former.  The consequences of
both alternative normalizations are considered in the present paper.

If CBP is strong enough, the \iso3{He} abundance in the stellar envelope
may reach an equilibrium value, where \iso3{He} destruction is balanced
by its creation via the \hbox{\it p\/-p\/}~chain.  This has been approximated
by using information from the parametric models combined with information
on the rates of the relevant reactions at the relevant temperatures~$T_P$
reached by the extra mixing.  For near-solar metallicity, CBP is not strong
enough on the RGB to attain the CBP-equilibrium \iso3{He} abundance in the
available time, and the combined effect of \iso3{He} creation and destruction
is fairly well constrained by the $1 \>M_\odot$, $Z = 0.02$ parametric models.
On the AGB (or on both RGB and AGB for low metallicity), the temperature~$T_P$
is relatively high, and the resulting
CBP-equilibrium \iso3{He} abundances are quite low (two or more orders of
magnitude below the initial stellar \iso3{He} abundances); if
equilibrium is reached, the remaining amount of \iso3{He} is negligible.

Figure~1 presents the predicted \iso3{He} enhancement/depletion factors
as a function of stellar mass for four different stellar metallicities.
The {\it heavy solid curve\/} shows the effect of first dredge-up on
the RGB (\markcite{Sackmann \& Boothroyd 1995a});
as noted above, first dredge-up leads to significant enhancement
of the surface \iso3{He} abundance in stars with masses
below $\sim 4\>M_\odot$.  For stars with masses above $\sim 2.3 \>M_\odot$,
first dredge-up takes place at the tip of the RGB, and there is no
subsequent CBP on the RGB (see \markcite{Boothroyd \& Sackmann 1995}).
However, in stars of lower mass, the onset of core helium burning is
delayed (due to the fact that the helium core is degenerate); these
stars have an extended RGB, and experience significant amounts of CBP\hbox{}.
The resulting \iso3{He} depletion (at the end of the
RGB) is indicated in Figure~1 by the {\it light solid curves\/}~``1c'';
note that they are normalized by the observations of post-first-dredge-up
RGB \iso{13}C production.

In stars of mass $\sim 4\>M_\odot$ and above, second dredge-up near the
beginning of the AGB causes a slight further depletion of~\iso3{He}
(\markcite{Sackmann \& Boothroyd 1995a}).
This is indicated in Figure~1 by the {\it heavy short-dashed curve}.
(Note that where this curve lies above curve~``1c'' in the $Z = 0.0001$
case, it should be ignored.  Above $\sim 7\>M_\odot$, second dredge-up
is interrupted by the ignition of carbon burning in the star's core;
the {\it dotted curve\/} shows the \iso3{He} abundance as of the time
of core carbon ignition.)  The \iso{18}O observations do not constrain
the time of occurence of CBP on the AGB; the most conservative assumption
is that it occurs throughout the AGB, as long as the hydrogen-shell is
burning and there is no composition gradient to oppose it, in analogy to
the RGB (see \markcite{Sackmann \& Boothroyd 1995a}).  A certain
amount of CBP is thus expected to occur
on the early AGB, during those periods when the hydrogen-shell is burning;
the result of this CBP for \iso3{He} is indicated by the {\it light
short-dashed curves} (``2c''~and~``2o'').  Curves~``2c'' show the effect
when the RGB \iso{13}C normalization of CBP is used, while curves~``2o'' show
the much stronger depletion that results from using the AGB \iso{18}O
normalization (these may be regarded as minimum and maximum depletion
estimates).

Eventually, periodic helium shell flashes (thermal pulses) begin to
take place on the AGB; their onset marks the end of the early AGB and
the beginning of the TP-AGB\hbox{}.  Stars of roughly $4 - 7\>M_\odot$
experience hot bottom burning during this stage of evolution, resulting
in depletion of~\iso3{He} by as much as 6 orders of magnitude
(\markcite{Sackmann \& Boothroyd 1992, 1995b}).  This is shown by the
{\it heavy long-dashed curves\/} in Figure~1.  For stars where hot bottom
burning does not take place, CBP is expected to
continue during the TP-AGB (interrupted only briefly by the periodic
helium shell flashes).  The TP-AGB is terminated when mass loss has
removed almost all the stellar envelope.  For stars of near-solar
metallicity, the initial--final mass relationship
(\markcite{Weidemann \& Koester 1983}; \markcite{Weidemann 1984})
allows one to estimate when this takes place, i.e., to estimate the
extent of the TP-AGB and thus the extent of CBP
there.  For low metallicities, the initial--final mass relationship is
sufficiently uncertain that no good estimate can be made.  Thus CBP
\iso3{He}-depletion on the TP-AGB is estimated only
in Figure~1a, as shown by the {\it light long-dashed curves\/} ``3c''
and~``3o'' (corresponding to the \iso{13}C and \iso{18}O normalizations,
respectively).

Note that \iso3{He} depletion factors for low-metallicity
stars are large enough in general that the lack of TP-AGB estimates for
these stars is of little importance.  Further, the final \iso3{He}
estimates of curves ``3c'' and~``3o'' may underestimate the {\it average\/}
\iso3{He} abundance in material returned to the interstellar medium due
to mass loss, which occurs throughout the AGB (though most strongly near
the end).  For this reason, and for consistency, the curves ``2c''
and~``2o'' were used as estimates of the \iso3{He} survival fraction
in material returned to the interstellar medium.
The curves~``2c'' in Figure~1 present a conservative estimate
of the \iso3{He} survival fraction resulting from dredge-up and CBP;
these may be supplemented by the hot bottom burning
curves above $\sim 4\>M_\odot$, and yield the ``low-CBP \iso3{He}-survival''
estimate.  The much lower \iso3{He} survival fractions of curves~``2o''
represent an extreme case, the ``high-CBP \iso3{He}-survival'' estimate.
These estimates of the ejected-matter \iso3{He}-survival fraction
as a function of stellar mass and metallicity are also tabulated in Table~1.

It should be noted that there is a significant uncertainty in CBP even on
the RGB\hbox{}.  There is some evidence suggesting that CBP might occur
episodically, rather than continuously throughout the relevant period
on the RGB (\markcite{de~la~Reza {\it et al.}\ 1995}); processing would
then take place at higher temperatures, and the observed amount of
\iso{13}C-creation might then correspond to a somewhat different amount
of \iso3{He}-destruction from that implied by the models of
\markcite{Sackmann \& Boothroyd (1995a)}.
Significant star-to-star variations are also possible, and it
is possible that some low mass stars experience no CBP at all, as suggested
by the high ratio $\rm \iso3{He}/H \sim 10^{-3}$ (by number) observed in
the planetary nebula NGC~3242 by \markcite{Rood, Bania, \& Wilson (1992)}.

\bigskip
\centerline{3. CHEMICAL EVOLUTION}
\medskip

Studies regarding the galactic evolution of D and $^3$He have a long
history, the details of which are described in recent works such as
\markcite{Vangioni-Flam {\it et al.}\ (1994)},
\markcite{Galli {\it et al.}\ 1995}, \markcite{Copi {\it et al.}\ (1995b)},
\markcite{Scully {\it et al.}\ (1995)}, and
\markcite{Dearborn {\it et al.}\ (1995)}.  By lowering the $^3$He survival
fractions in a parameterized fashion,
these works attempt to include in the chemical evolution the
effects of $^3$He destruction in low mass stars.  Here, we wish
to couple our {\it detailed} estimates of $^3$He destruction as listed
in Table~1, into models of D and $^3$He chemical evolution.
Our primary goal is to establish the increase in the inferred
primordial (D$+^3$He)/H ratio which can reasonably be expected,
given that extra mixing processes on the RGB may be occurring.  We
stress that it is not our role to exhaustively
investigate D$+^3$He galactic chemical evolution.  Such a study
would involve a myriad of possibilities, such as the
infall and mixing of the interstellar gas.
We merely wish to explore the potential effects of extra RGB mixing
on typical chemical evolution models, and to determine if such mixing is
an important issue for cosmological constraints.
To this end we will adopt the chemical evolution
models of \markcite{Vangioni-Flam {\it et al.}\ (1994)} as the basis
of our study.  Vangioni-Flam {\it et al.}\
highlight the possible chemical evolution models which can accommodate not
only the pre-solar and present D and $^3$He data, but also a variety
of other constraints such as the metallicity and gas evolution of the galaxy.
Use of these models will at least allow us to gauge the importance of
our new $^3$He survival fractions.

The differential equations describing the evolution of the
gas mass fraction $\sigma_{\!g}$, and the element mass fraction $X_i$ can be
written (e.g., \markcite{Tinsley 1980})
$$
{d\sigma_{\!g}\over dt}=   \ { \int^{m_{up}}_{m(t)} (m-m_r)
 \Psi(t^f_m) \Phi(m) dm
-\Psi(t) }   \ \ \ ,
\eqno(1a)
$$
and
$$
{d(\sigma_{\!g}\,X_i)\over dt}=   \ { \int^{m_{up}}_{m(t)} m
 \Psi(t^f_m) \Phi(m)Y_i(t^f_m) dm
-X_i\Psi(t) }   \ \ \ .
\eqno(1b)
$$
Here $t^f_m$ is the formation time at which a star of mass~$m$ is
returning gas back to the interstellar medium at the current time~$t$.
$\Psi$~is the star formation rate (SFR), which we assume to be given by
$\Psi=0.25\,\sigma_{\!g}$.
$Y_i$~is the elemental stellar yield, written as  the mass of element~$i$
in the ejecta divided by the initial mass of the star
from which it was ejected.
The lower limit of integration $m(t)$ is the minimum stellar mass which
can be returning gas to the interstellar medium at time~$t$.
For the remnant mass function~$m_r$, the relations
of \markcite{Iben and Tutukov (1984)} are adopted; and for the stellar
lifetimes the relations of \markcite{Scalo (1986)} are adopted.
For the initial mass function (IMF)~$\Phi(m)$, we
assume a power law $\Phi(m) \propto m^{-(1+x)}$,
normalized through the relation
$$
\int^{m_{up}}_{m_{low}} m \Phi(m)dm=1 \ \ \ .
\eqno(2)
$$
The upper and lower mass limits are taken to be $m_{up}=100 \>M_\odot$ and
$m_{low}=0.4 \>M_\odot$, respectively, and we adopt $x=1.7$. To this model
we couple in the  metallicity dependent $^3$He survival fractions of Table~1
by adopting a metallicity evolution compatible with presently available
data (\markcite{Edvardsson {\it et al.}\ 1993}).
For more massive stars ($m > 9 \>M_\odot$) we use
a metallicity independent value of $g_3=0.25$
(e.g., \markcite{Dearborn {\it et al.}\ 1986}),
although the results are quite insensitive to this value.

Assuming the low-CBP survival fractions (i.e., minimal ${}^3$He destruction),
Figure~2a displays the results of our calculations for various values of the
primordial (D$+^3$He)/H ratio. Here we assume a galactic age of 14~Gyr. The
data point corresponds to the presolar (D$+^3$He)/H ratio as reported by
\markcite{Geiss (1993)}, with the error bars indicating the uncertainty at the
$2\sigma$ level. We assume that the presolar value is representative
of the galactic value (local anomalous $^3$He values offer
another means of achieving compatibility, see for example
\markcite{Olive {\it et al.}\ [1995]}).
That the (D$+^3$He)/H curve pass through this data point
is the key test regarding the viability of our initial conditions.
In order to satisfy the presolar data we can see from Figure~2a, that the
initial value of the (D$+^3$He)/H ratio cannot be greater than
$1 \times 10^{-4}$. This is essentially the same limit utilized in current
studies of SBBN, and resembles the results of model $I_{a,e}$ of
\markcite{Vangioni-Flam {\it et al.}\ (1994)} who found that an unexplained
reduction of $g_3$ ($g_3=0.5$, 0.3, 0.3 for $M= 1$, 2, $3 \>M_\odot$)
was necessary for consistency with SBBN.

{}From the above calculations then, we find
that extra mixing on the RGB can indeed lead to lower values of $g_3$, but
that the extent
of this effect only allows for a consistent explanation of the
galactic evolution of $^3$He. More importantly, it does not allow for
the lower constraint on $\Omega_b$ to be altered.

We have repeated our calculation for the high-CBP survival fractions (i.e.,
more extreme ${}^3$He destruction),
the results of which are shown in Figure~2b. It can be seen that these revised
survival fractions can influence the results
in only in a relatively small way.  This is largely a consequence of the
fact that even for extremely low $g_3$ values,  D$+^3$He cannot be greatly
reduced beyond that shown so long as a residual component (say~5\%)
of the initial gas remains unprocessed.
In order to satisfy the presolar data we see from Figure~2b that the
initial value of the (D$+^3$He)/H ratio cannot be greater than
$1.2\times 10^{-4}$.

We have explored numerous variations in our  chemical
evolution models such as different mass loss
rates, SFR laws and IMF's. However, if consistency with all the
observable data is to be maintained, the inferred primordial
(D$+^3$He)/H ratio cannot be increased above
(D$+^3$He)/H$\approx 1.5 \times 10^{-4}$
(note that this bound is marginally consistent with the value of
$\rm D/H \approx 2.5 \times 10^{-4}$ observed in a high-redshift quasar
absorber by \markcite{Songaila {\it et al.}\ [1994]} and
\markcite{Carswell {\it et al.}\ [1994]}).  It is a challenge to model builders
to create a reasonable and self-consistent
model of galactic chemical evolution which can significantly
circumvent this bound.

To put a small change of $0.5 \times 10^{-4}$ in the
upper limit to the inferred primordial
(D$+^3$He)/H ratio into perspective, we note that it lowers the
lower bound on the baryonic density only minimally
to $\Omega_b\approx 0.008$, and
provides room for additional statistical weight
at the epoch of nucleosynthesis of roughly $0.5$ equivalent neutrino families.
Coupled with the other uncertainties regarding SBBN
(e.g., \markcite{Copi {\it et al.}\ 1995a}),
this would allow for the equivalent of approximately one
new neutrino family in the early universe.

\bigskip
\centerline{4. CONCLUSION}
\medskip

We have investigated in some detail the effect of extra mixing
on the RGB as a source of stellar $^3$He destruction.  Through its effects
on chemical evolution, we determine the consequences of such $^3$He
destruction for the primordial bound on
D$+^3$He.  We find that although a consistent picture of galactic evolution
for D$+^3$He can emerge, the allowed increase in the primordial
(D$+^3$He)/H ratio is small, and its impact on the lower limit
to $\Omega_b$ is minimal.  In the extreme $^3$He-destruction case, a small
increase in the relativistic degrees of freedom in the early universe
can be accommodated.

\bigskip

We are indebted to I.-J. Sackmann for encouragement and helpful discussions,
and wish to thank
S.~D. Tremaine and P.~G. Martin for the support provided by the
Canadian Institute for Theoretical Astrophysics.  This work was supported in
part by a grant from the Natural Sciences and Engineering Research Council
of Canada.

\vfill\eject

\centerline{REFERENCES}
\medskip

{\parskip=0pt

\par\noindent\hang
 Boothroyd,~A.~I., \& Sackmann,~I.-J. 1995, ApJ, submitted;
  preprint astro-ph/9512121
\par\noindent\hang
 Boothroyd,~A.~I., Sackmann,~I.-J., \& Wasserburg,~G.~J. 1995, ApJ, 442, L21
  \ \ (BSW95)
\par\noindent\hang
 Carswell,~R.~F., Rauch,~M., Weymann,~R.~J., Cooke,~A.~J., \& Webb,~J.~K. 1994,
  MNRAS, 268, L1
\par\noindent\hang
 Charbonnel,~C. 1994, A\&A, 282, 811
\par\noindent\hang
 ---. 1995, preprint astro-ph/9511080
\par\noindent\hang
 Copi, C. J., Schramm, D. N., \& Turner, M. S. 1995a, preprint astro-ph/9508029
\par\noindent\hang
 ---. 1995b, preprint astro-ph/9506094
\par\noindent\hang
 Dearborn,~D.~S.~P. 1992, Phys.\ Reports, 210, 367
\par\noindent\hang
 Dearborn,~D., Eggleton,~P.~P., \& Schramm,~D.~N. 1976, ApJ, 203, 455
\par\noindent\hang
 Dearborn, D. S. P., Schramm, D. N., \& Steigman, G. 1986, ApJ, 302, 35
\par\noindent\hang
 Dearborn, D. S. P., Steigman, G., \& Tosi, M. 1995, preprint OSU-TA-16/95
\par\noindent\hang
 de~la~Reza,~R., Drake,~N.~A., \& da~Silva,~L. 1995, ApJ (Lett), submitted
\par\noindent\hang
 Denissenkov,~P.~A., \& Weiss,~A. 1995, A\&A, submitted
\par\noindent\hang
 Edvardsson, B., Anderson, J., Gustafson, B., Lambert, D. L., Nissen, P. E.,
  \& Tomkin, J. 1993, A\&A, 275, 101
\par\noindent\hang
 Galli, D., Palla, F., Ferrini, F., \& Penco, U. 1995, ApJ, 443, 536
\par\noindent\hang
 Galli,~D., Palla,~F., Straniero,~O., \& Ferrini,~F. 1994, ApJ, 432, L101
\par\noindent\hang
 Geiss, J. 1993, in {\it Origin and Distribution of the Elements}, ed.\
  N. Prantzos, E. Vangioni-Flam, \& M. Cass\'e (Cambridge; Cambridge
  University Press), 89
\par\noindent\hang
 Genova,~F., \& Schatzmann,~E. 1979, A\&A, 78, 323
\par\noindent\hang
 Gilroy,~K.~K. 1989, ApJ, 347, 835
\par\noindent\hang
 Gilroy,~K.~K., \& Brown,~J.~A. 1991, ApJ, 371, 578
\par\noindent\hang
 Hogan, C. 1995, ApJ, 441, L17
\par\noindent\hang
 Iben,~I.,~Jr. 1967, ApJ, 147, 624
\par\noindent\hang
 Iben, I., Jr., \& Tutukov,~A. 1984, ApJS,  54, 335
\par\noindent\hang
 Kernan, P., \& Krauss, L. 1994, Phys.\ Rev.\ Lett., 72, 3309
\par\noindent\hang
 Levshakov,~S.~A., \& Takahara,~F. 1995, preprint
\par\noindent\hang
 Olive, K. A., Rood, R. T., Schramm, D. N., Truran, J. W., \&
  Vangioni-Flam, E. 1995, ApJ, 444, 680
\par\noindent\hang
 Rood,~R.~T., Bania,~T.~M., \& Wilson,~T.~L. 1984, ApJ, 280, 629
\par\noindent\hang
 ---. 1992, Nature, 355, 618
\par\noindent\hang
 Sackmann,~I.-J., \& Boothroyd,~A.~I. 1992, ApJ, 392, L71
\par\noindent\hang
 ---. 1995a, ApJ, submitted; preprint astro-ph/9512122
\par\noindent\hang
 ---. 1995b, in preparation
\par\noindent\hang
 Sackmann,~I.-J., Boothroyd,~A.~I., \& Kraemer,~K.~E. 1993, ApJ, 418, 457
\par\noindent\hang
 Scalo, J. M. 1986, Fund.\ Cosm.\ Phys., 11, 1
\par\noindent\hang
 Scully, S. T., Cass\'e, M., Olive, K. A., Schramm, D. N., Truran, J. W., \&
  Vangioni-Flam, E. 1995, preprint astro-ph/9508086
\par\noindent\hang
 Smith,~G.~H., \& Tout,~C.~A. 1992, MNRAS, 256, 449
\par\noindent\hang
 Smith, M. S., Kawano, L. H., \& Malaney, R. A. 1993, ApJS, 85, 219
\par\noindent\hang
 Songaila,~A., Cowie,~L.~L., Hogan,~C.~J., \& Rugers,~M. 1994, Nature, 368, 599
\par\noindent\hang
 Tinsley, B. M. 1980, Fund.\ Cosm.\ Phys., 5, 287
\par\noindent\hang
 Tytler,~D., \& Fan,~X. 1994, BAAS, 26, 1424
\par\noindent\hang
 Vangioni-Flam, E., Olive, K. A., \& Prantzos, N. 1994, ApJ, 427, 618
\par\noindent\hang
 Walker, T. P., Steigman, G., Schramm, D. N., Olive, K. A, \& Kang, H. 1991,
  ApJ, 376, 51
\par\noindent\hang
 Wasserburg,~G.~J., Boothroyd,~A.~I., \& Sackmann,~I.-J. 1995, ApJ, 447, L37
  \ \ (WBS95)
\par\noindent\hang
 Weidemann,~V. 1984, A\&A, 134, L1
\par\noindent\hang
 Weidemann,~V., \& Koester,~D. 1983, A\&A, 121, 77
\par\noindent\hang
 Yang, J., Turner, M. S., Steigman, G., Schramm, D. N., \& Olive, K. A. 1994,
  ApJ, 281, 493

}   

\vfill
\eject

\centerline{TABLE 1}
\centerline{\iso3{He} Survival Fractions$^{\,a}$ in Ejected Stellar
 Material From Stars of Mass $M$}
\smallskip

{
\tabskip=0 pt \parskip=0 pt
\halign to \hsize{
 \hfil $#$\hfil\tabskip=0.4 em plus 1 fil& \hfil $#$\hfil& \hfil $#$\hfil&
 \hfil $#$\hfil& \hfil $#$\hfil& \hfil $#$\hfil& \hfil $#$\hfil&
 \hfil $#$\hfil& \hfil $#$\hfil\tabskip=0 pt\cr
\noalign{\medskip\hrule\smallskip\hrule\medskip}
 & \multispan2\hfill$Z=0.02$\hfill& \multispan2\hfill$Z=0.007$\hfill&
 \multispan2\hfill$Z=0.001$\hfill& \multispan2\hfill$Z=0.0001$\hfill\cr
\noalign{\vskip -9pt}
M& \multispan2\hrulefill& \multispan2\hrulefill& \multispan2\hrulefill&
 \multispan2\hrulefill\cr
\noalign{\vskip -3pt}
(M_\odot)& \hbox{low-CBP}& \hbox{high-CBP}& \hbox{low-CBP}& \hbox{high-CBP}&
 \hbox{low-CBP}& \hbox{high-CBP}& \hbox{low-CBP}& \hbox{high-CBP}\cr
\noalign{\medskip\hrule\medskip}
0.85&      &       &       &        & 0.0246&
 1.91(-4)\rlap{\hbox{$^{\,b}$}}& 0.00184& 4.41(-5)\cr
0.90&      &       & 0.162 & 0.00125& 0.0291& 1.91(-4)& 0.00268& 4.41(-5)\cr
0.95&      &       & 0.185 & 0.00227& 0.0337& 1.91(-4)& 0.00377& 4.41(-5)\cr
1.00& 0.463& 0.0124& 0.209 & 0.00409& 0.0384& 1.91(-4)& 0.00505& 4.41(-5)\cr
1.10& 0.555& 0.0261& 0.254 & 0.0108 & 0.0486& 1.97(-4)& 0.00789& 4.41(-5)\cr
1.20& 0.626& 0.0456& 0.304 & 0.0221 & 0.0596& 2.73(-4)& 0.0107 & 4.41(-5)\cr
1.35& 0.737& 0.0691& 0.375 & 0.0371 & 0.0755& 6.42(-4)& 0.0116 & 4.41(-5)\cr
1.50& 0.832& 0.0966& 0.446 & 0.0556 & 0.0936& 0.00167 & 0.0157 & 4.41(-5)\cr
1.65& 0.903& 0.124 & 0.513 & 0.0749 & 0.115 & 0.00350 & 0.0403 & 4.41(-5)\cr
1.80& 0.977& 0.154 & 0.596 & 0.0971 & 0.144 & 0.00643 & 0.195  & 4.42(-5)\cr
2.00& 1.106& 0.199 & 0.789 & 0.134  & 0.261 & 0.0138  & 0.213  & 4.43(-5)\cr
2.25& 1.371& 0.266 & 1.650 & 0.206  & 1.019 & 0.0308  & 0.229  & 4.45(-5)\cr
2.50& 1.439& 0.302 & 1.496 & 0.283  & 1.007 & 0.0341  & 0.245  & 4.52(-5)\cr
2.75& 1.310& 0.397 & 1.354 & 0.326  & 1.021 & 0.0423  & 0.288  & 5.29(-5)\cr
3.00& 1.187& 0.481 & 1.235 & 0.346  & 1.051 & 0.0747  & 0.373  & 1.99(-4)\cr
3.25& 1.090& 0.512 & 1.146 & 0.357  & 1.110 & 0.218   & 0.580  & 0.00588 \cr
3.50& 1.014& 0.535 & 1.084 & 0.411  & 1.019 & 0.247   & 0.1    & 0.00477 \cr
3.75& 0.957& 0.551 & 0.316 & 0.316  & 0.1   & 0.1     & 0.01   & 0.00790 \cr
4.00& 0.316& 0.316 & 0.1   &   0.1  & 0.01  & 0.01    & 0.001  & 0.001   \cr
4.50& 0.1  & 0.1   & 0.01  &   0.01 & 0.001 & 0.001   & 1.(-6) & 1.(-6)  \cr
5.00& 0.01 & 0.01  & 0.001 &  0.001 & 1.(-6)& 1.(-6)  & 1.(-6) & 1.(-6)  \cr
5.50& 0.001& 0.001 & 1.(-4)& 1.(-4) & 1.(-6)& 1.(-6)  & 1.(-6) & 1.(-6)  \cr
6.00& 0.001& 0.001 & 1.(-4)& 1.(-4) & 1.(-6)& 1.(-6)  & 1.(-6) & 1.(-6)  \cr
6.50& 0.001& 0.001 & 1.(-4)& 1.(-4) & 0.589 & 0.589   & 0.505  & 0.505   \cr
7.00& 0.001& 0.001 & 0.607 & 0.607  & 0.563 & 0.563   & 0.477  & 0.477   \cr
7.50& 0.568& 0.568 & 0.584 & 0.584  & 0.560 & 0.560   & 0.495  & 0.495   \cr
8.00& 0.547& 0.547 & 0.581 & 0.581  & 0.600 & 0.600   & 0.501  & 0.501   \cr
8.50& 0.522& 0.522 & 0.631 & 0.631  & 0.588 & 0.588   & 0.484  & 0.484   \cr
9.00& 0.538& 0.538 & 0.622 & 0.622  & 0.573 & 0.573   & 0.471  & 0.471   \cr
\noalign{\medskip\hrule\smallskip\hrule}
}}
\medskip

$^a$ The ratio of the \iso3{He} abundance (by mass fraction) in ejected
material to its initial (input) stellar abundance (by mass fraction)
$\iso3{He}_{\rm init}$; ``low-CBP'' corresponds to curves~``2c'' of Fig.~1,
and ``high-CBP'' to curves~``2o'' of Fig.~1, in both cases supplemented by
the effect of hot bottom burning in the relevant mass range.

$^b$ Power of ten notation: $1.91(-4) \equiv 1.91 \times 10^{-4}$.

\vfill
\eject

\centerline{FIGURE CAPTIONS}
\medskip

Fig.~1.---Log of the \iso3{He} survival fraction in the stellar envelope,
i.e., the ratio of the \iso3{He} abundance (by mass fraction) in the stellar
envelope to its initial (input) stellar abundance (by mass fraction)
$\iso3{He}_{\rm init}$.  Effects of standard dredge-up and (normalized)
cool bottom processing estimates are shown, as a
function of stellar mass (see text for meaning of curves).
(a)~For solar ($Z = 0.02$) and near-solar ($Z = 0.007$) metallicity.
(b)~For low (Population~II) metallicities ($Z = 0.001$ and~0.0001).
\medskip

Fig.~2a.---The galactic evolution of (D$+^3$He)/H for low-CBP case, assuming
initial values of 12, 10, 9 $\times 10^{-5}$ (top to bottom).
\medskip

Fig.~2b.---The galactic evolution of (D$+^3$He)/H for high-CBP case, assuming
initial values of 15, 13, 12, 11, 10 $\times 10^{-5}$ (top to bottom).

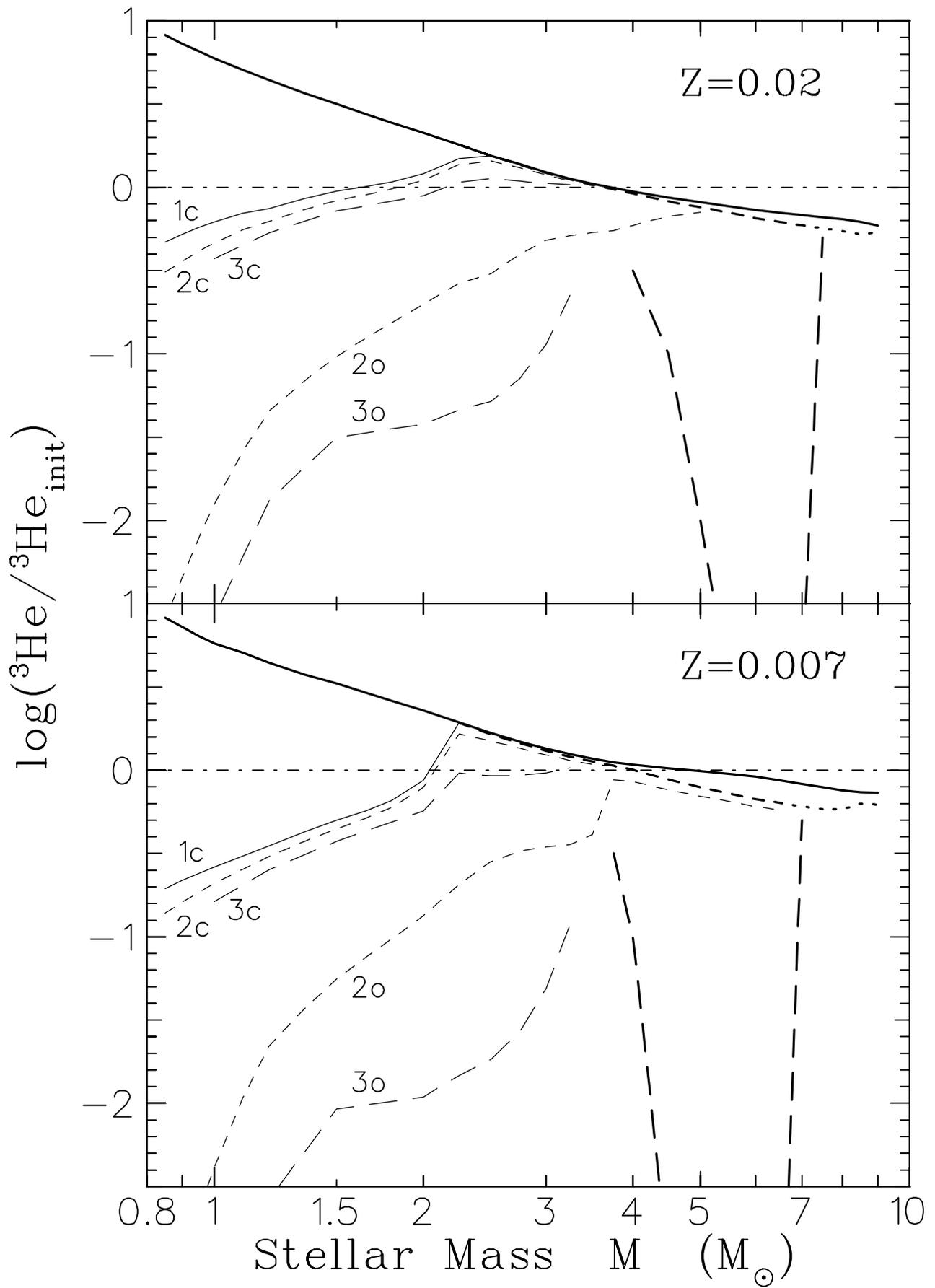
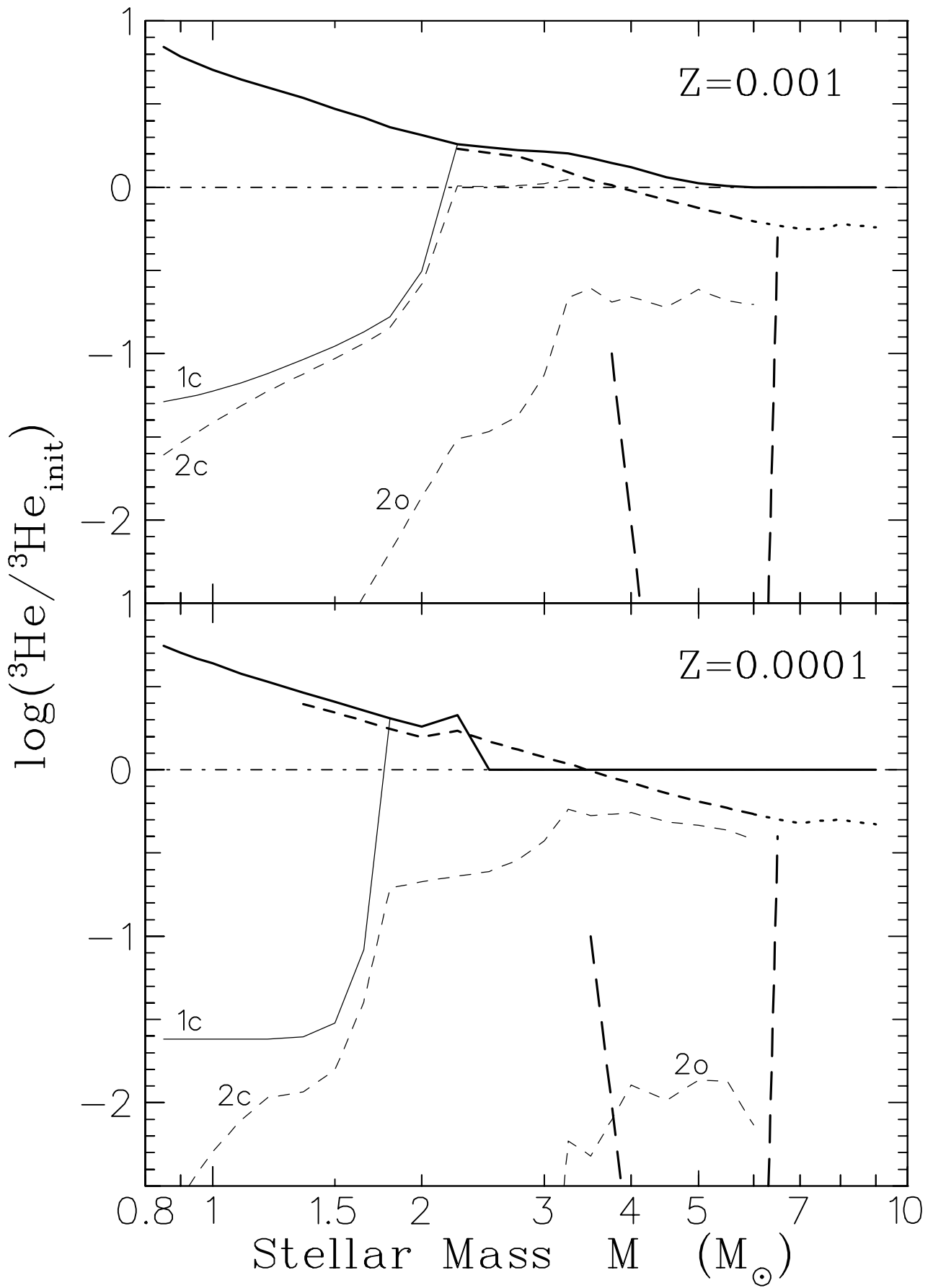
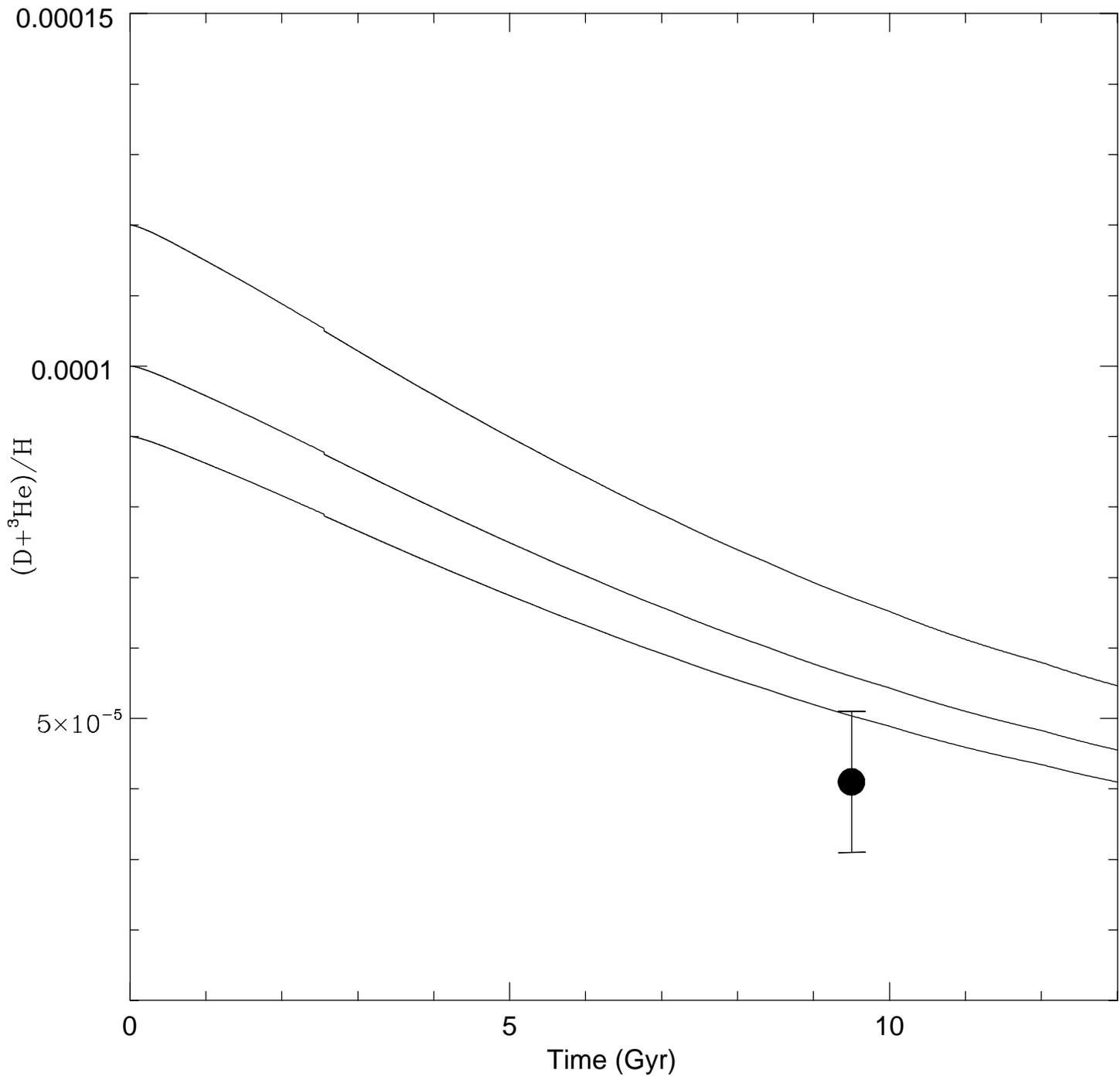
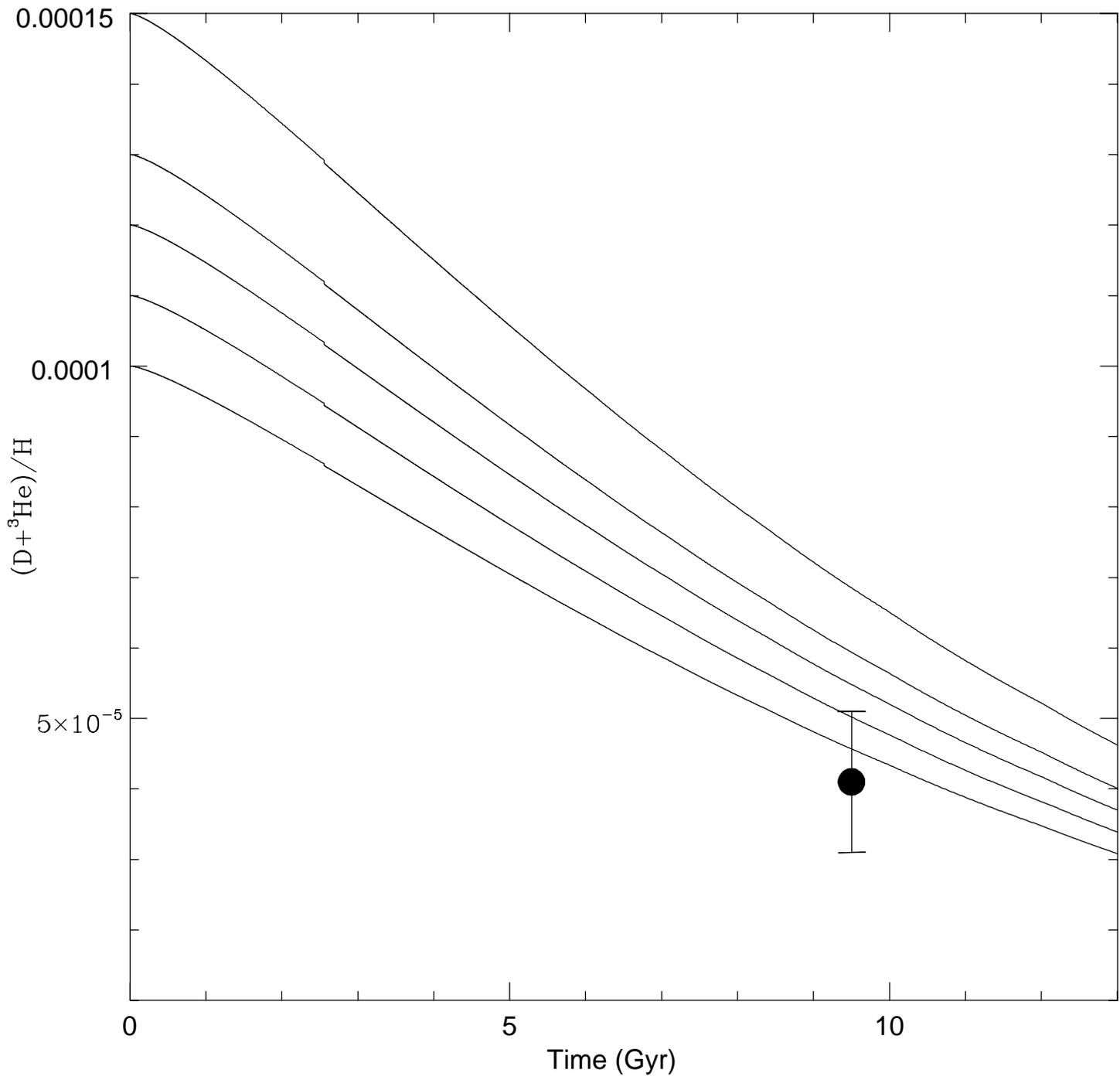
\bye